
\input harvmac

\def\inbar{\,\vrule height1.5ex width.4pt depth0pt}
\def\IB{\relax{\rm I\kern-.18em B}}
\def\IC{\relax\hbox{$\inbar\kern-.3em{\rm C}$}}
\def\ID{\relax{\rm I\kern-.18em D}}
\def\IE{\relax{\rm I\kern-.18em E}}
\def\IF{\relax{\rm I\kern-.18em F}}
\def\IG{\relax\hbox{$\inbar\kern-.3em{\rm G}$}}
\def\IH{\relax{\rm I\kern-.18em H}}
\def\II{\relax{\rm I\kern-.18em I}}
\def\IK{\relax{\rm I\kern-.18em K}}
\def\IL{\relax{\rm I\kern-.18em L}}
\def\IM{\relax{\rm I\kern-.18em M}}
\def\IN{\relax{\rm I\kern-.18em N}}
\def\IO{\relax\hbox{$\inbar\kern-.3em{\rm O}$}}
\def\IP{\relax{\rm I\kern-.18em P}}
\def\IQ{\relax\hbox{$\inbar\kern-.3em{\rm Q}$}}
\def\IR{\relax{\rm I\kern-.18em R}}
\font\cmss=cmss10 \font\cmsss=cmss10 at 7pt
\def\IZ{\relax\ifmmode\lrefhchoice
{\hbox{\cmss Z\kern-.4em Z}}{\hbox{\cmss Z\kern-.4em Z}}
{\lower.9pt\hbox{\cmsss Z\kern-.4em Z}}
{\lower1.2pt\hbox{\cmsss Z\kern-.4em Z}}\else{\cmss Z\kern-.4em Z}\fi}
\def\IGa{\relax\hbox{${\rm I}\kern-.18em\Gamma$}}
\def\IPi{\relax\hbox{${\rm I}\kern-.18em\Pi$}}
\def\ITh{\relax\hbox{$\inbar\kern-.3em\Theta$}}
\def\IOm{\relax\hbox{$\inbar\kern-3.00pt\Omega$}}

\def\np{Nucl. Phys. }
\def\pl{Phys. Lett. }
\def\pr{Phys. Rev. }
\def\prl{Phys. Rev. Lett. }

\def \ab{\bar{a}}

\def \l{\ell}

\def \cosh{{\rm cosh}}

\def\inbar{\,\vrule height1.5ex width.4pt depth0pt}
\def\IB{\relax{\rm I\kern-.18em B}}
\def\IC{\relax\hbox{$\inbar\kern-.3em{\rm C}$}}
\def\IP{\relax{\rm I\kern-.18em P}}
\def\IR{\relax{\rm I\kern-.18em R}}
\def\inbar{\,\vrule height1.5ex width.4pt depth0pt}
\def\IB{\relax{\rm I\kern-.18em B}}
\def\IC{\relax\hbox{$\inbar\kern-.3em{\rm C}$}}
\def\ID{\relax{\rm I\kern-.18em D}}
\def\IE{\relax{\rm I\kern-.18em E}}
\def\IF{\relax{\rm I\kern-.18em F}}
\def\IG{\relax\hbox{$\inbar\kern-.3em{\rm G}$}}
\def\IH{\relax{\rm I\kern-.18em H}}
\def\II{\relax{\rm I\kern-.18em I}}
\def\IK{\relax{\rm I\kern-.18em K}}
\def\IL{\relax{\rm I\kern-.18em L}}
\def\IM{\relax{\rm I\kern-.18em M}}
\def\IN{\relax{\rm I\kern-.18em N}}
\def\IO{\relax\hbox{$\inbar\kern-.3em{\rm O}$}}
\def\IP{\relax{\rm I\kern-.18em P}}
\def\IQ{\relax\hbox{$\inbar\kern-.3em{\rm Q}$}}
\def\IR{\relax{\rm I\kern-.18em R}}
\font\cmss=cmss10 \font\cmsss=cmss10 at 7pt
\def\IZ{\relax\ifmmode\mathchoice
{\hbox{\cmss Z\kern-.4em Z}}{\hbox{\cmss Z\kern-.4em Z}}
{\lower.9pt\hbox{\cmsss Z\kern-.4em Z}}
{\lower1.2pt\hbox{\cmsss Z\kern-.4em Z}}\else{\cmss Z\kern-.4em Z}\fi}
\def\IGa{\relax\hbox{${\rm I}\kern-.18em\Gamma$}}
\def\IPi{\relax\hbox{${\rm I}\kern-.18em\Pi$}}
\def\ITh{\relax\hbox{$\inbar\kern-.3em\Theta$}}
\def\IOm{\relax\hbox{$\inbar\kern-3.00pt\Omega$}}

\def\l {\ell }

\def\Tr {\rm Tr}

\def\pl{Phys. Lett. B }
\def\np{Nucl. Phys.B }
\def\pr{Phys. Rev. }
\def\prl{Phys. Rev. Lett. }

\lref\Iml{I.K. Kostov, Phys. Lett. B 266 (1991) 42}
\lref\ks{I. Kostov and M. Staudacher, to be published}
\lref\Icar{I.Kostov, ``Strings embedded in Dynkin diagrams'',  Lecture
given at the Cargese meeting, Saclay preprint
SPhT/90-133.}
\lref\Mig{A.A. Migdal, Phys. Rep. 102 (1983) 199}
\lref\bkz{E. Br\'ezin, V. Kazakov, and Al. B. Zamolodchikov,
\Nucl. Phys. B338 (1990) 673}
\lref\Iade{I. Kostov, Nucl. Phys. B 326 (1989)583.}
\lref\bkkm{D. Boulatov, V. Kazakov, I. Kostov, and A.A. Migdal, Nucl. Phys. B
275 [FS 17] (1986) 641}
\lref\Inonr{I. Kostov, \pl 266 (1991) 317.}
\lref\bpz{A. Belavin, A. Polyakov, and A. Zamolodchikov, Nucl. Phys.
B 241 (1984) 333}
\lref\df{V.Dotsenko and V. Fateev, Nucl. Phys. B 240 (1984) 312}
\lref\Imult{I. Kostov,
\pl 266 (1991) 42.}
\lref\ajm{J. Ambjorn, J. Jurkiewicz and Yu. Makeenko, \pl 251 (1990)517}
\lref\mat{V. Kazakov, Phys. Lett. 150B (1985) 282;
F. David, Nucl. Phys. B 257 (1985) 45;
V. Kazakov, I. Kostov and A.A. Migdal, Phys. Lett. 157B (1985),295;
J. Ambjorn, B. Durhuus, and J. Fr\"ohlich, Nucl. Phys. B 257 (1985) 433}
\lref\brkz{E. Br\'ezin and V. Kazakov, Phys. Lett. 236B (1990) 144.}
\lref\Ion{I. Kostov, Mod. Phys. Lett. A4 (1989) 217}
\lref\dglsh{M. Douglas and S. Shenker, Nucl. Phys. B 335 (1990) 635.}
\lref\abf{G. Andrews, R. Baxter , and P. Forrester, J. Stat. Phys. 35
 (1984) 35}
\lref\bax{R. Baxter, Exactly Solvable Models in Statistical Mechanics
(Academic Press, London,1982)}
\lref\grmg{D. Gross and A. Migdal, Phys. Rev. Lett. 64 (1990) 127.}
\lref\mike{M. Douglas, Phys. Lett. 238B (1990) 176.}
\lref\pol{A. Polyakov, Phys. Lett. 103 B (1981) 207, 211.}
\lref\kpz{V. Knizhnik, A. Polyakov and A. Zamolodchikov, Mod. Phys. Lett.
A3 (1988) 819.}
\lref\ddk{F. David, Mod. Phys. Lett. A3 (1988) 1651; J. Distler and
H. Kawai, Nucl. Phys. B 321 (1989) 509}
\lref\bk{M. Bershadski, I. Klebanov, Nucl. Phys. B 360 (1991) 559. }
\lref\mss{G.Moore, N.Seiberg and M. Staudacher, Nucl. Phys. B 362 (1991)665.}
\lref\polci{J. Polchinski, Nucl. Phys. B346 (1990) 253 }
\lref\stau{M. Staudacher,Nucl. Phys. B 336 (1990) 349.}
\lref\gm{D. Gross and A. Migdal, Nucl. Phys. B340 (1990) 333}
\lref\David{F. David, Mod. Phys. Lett. A, No.13 (1990) 1019}
\lref\bax{R. J. Baxter, Exactly solved Models in Statistical Mechanics
(Academic Press, 1982) }

\lref\wdw{J.B. Hartle and S.W. Hawking, Phys. Rev. {\bf D28} (1983)
2960.}
\lref\teit{C. Teitelboim, Phys. Rev. Lett. {\bf 50} (1983) 705.}
\lref\mat{J. Ambj{\o}rn, B. Durhuus, J. Fr\"ohlich, Nucl. Phys.
{\bf B257} (1985) 433; F. David, Nucl. Phys. {\bf B257} (1985) 45;
V. Kazakov, Phys. Lett. {\bf 150B} (1985) 282;
V. Kazakov, I. Kostov, A. Migdal, Phys. Lett. {\bf 157B} (1985) 295.}
\lref\brkz{E. Br\'ezin and V. Kazakov, Phys. Lett. {\bf 236B} (1990) 144.}
\lref\dglsh{M. Douglas and S. Shenker, Nucl. Phys. {\bf B335} (1990) 635.}
\lref\grmg{D. Gross and A. Migdal, Phys. Rev. Lett. {64} (1990) 127.}
\lref\mike{M. Douglas, Phys. Lett 238B (1990) 176.}
\lref\pol{A. Polyakov, Phys. Lett. {\bf 103B} (1981) 207, 211.}
\lref\kpz{V. Knizhnik, A. Polyakov and A. Zamolodchikov, Mod. Phys. Lett.
{\bf A3} (1988) 819.}
\lref\ddk{F. David, Mod. Phys. Lett. {\bf A3} (1988) 1651;
J. Distler and H. Kawai, Nucl. Phys. {\bf B321} (1989) 509.}
\lref\other{
J. Polchinski,``Remarks on the Liouville Field Theory,'' Texas preprint
UTTG--19--90}
\lref\otheri{J. Polchinski, Nucl. Phys. {\bf B324} (1989) 123;
J. Polchinski, Nucl. Phys. {\bf B346} (1990) 253;
J. Polchinski,``Ward Identities in Two Dimensional
Gravity,'' Texas preprint UTTG--39--90.}
\lref\nati{N. Seiberg,``Notes on Quantum Liouville Theory and Quantum
Gravity,'' Rutgers preprint RU--90--29, to appear in the proceedings of
the 1990 Yukawa International Seminar, and in the proceedings of the
Carg\`ese Workshop on Random Surfaces.}
\lref\gou{A. Gupta, S. Trivedi and M. Wise, Nucl. Phys. {\bf B340} (1990)
475; M. Bershadsky and I. Klebanov, \prl {\bf 65}(1990)3088;
M. Goulian and M. Li, ``Correlation Functions in Liouville
Theory,'' Santa Barbara preprint UCSBTH--90--61;
P. Di Francesco and D. Kutasov, ``Correlation Functions in 2D String
Theory,'' Princeton preprint PUPT--1237.}
\lref\matt{M. Staudacher, Nucl. Phys. {\bf B336} (1990) 349.}
\lref\bdks{E. Br\'ezin, M. Douglas, V. Kazakov and S. Shenker, Phys.
Lett. B237 (1990) 43;
D. Gross and M. Migdal, Phys. Rev. Lett. 64 (1990) 717;
\v C. Crnkovi\'c, P. Ginsparg and G. Moore, Phys. Lett.
237B (1990) 196.}
\lref\sasha{A.B. Zamolodchikov, unpublished, and private communications.}
\lref\bds{T. Banks, N. Seiberg and S. Shenker, unpublished.}
\lref\vv{E. Verlinde and H. Verlinde, Nucl. Phys. {\bf B348} (1991) 457.}
\lref\dvv{R. Dijkgraaf, E. Verlinde and H. Verlinde, Nucl. Phys.
{\bf B348} (1991) 435; R. Dijkgraaf, E. Verlinde and H. Verlinde,
``Notes on Topological String Theory and 2D Quantum Gravity,''
Princeton preprint PUPT--1217, presented at the Carg\`ese Workshop.}
\lref\fkn{M. Fukuma, H. Kawai and R. Nakayama,``Continuum
Schwinger--Dyson Equations and Universal Structures in Two--Dimensional
Quantum Gravity,'' Tokyo preprint UT--562.}
\lref\kuta{D. Kutasov, Phys. Lett. {\bf 220B} (1989) 153.}
\lref\wegner{find reference}
\lref\emilpaper{E. Martinec, G. Moore and N. Seiberg, Rutgers preprint
RU--91--14.}
\lref\boulkz{V. Kazakov, Phys. Lett. {\bf 119A} (1986) 140;
D. Boulatov and V. Kazakov, Phys. Lett. {\bf 186B} (1987) 379.}
\lref\kutdi{P. Di Francesco and D. Kutasov, Nucl. Phys. B342 (1990)
589; P. Di Francesco and D. Kutasov,``Integrable Models of
Two--Dimensional Quantum Gravity,''Princeton preprint PUPT--1206 (1990),
presented at the Carg\`ese Workshop.}
\lref\bpz{A. Belavin, A. Polyakov and A. Zamolodchikov, Nucl. Phys.
{\bf B241} (1984) 333.}
\lref\mtloo{G. Moore, Phys. Lett. {\bf 176B} (1986) 369.}
\lref\bdss{T. Banks, M. Douglas, N. Seiberg and S. Shenker, Phys. Lett.
238B (1990) 279.}
\lref\nonline{S. Wadia, Phys. Rev. {\bf D24} (1981) 970;
A. Migdal, Phys. Rep. {\bf 102} (1983) 199;
V. Kazakov, Mod. Phys. Lett. {\bf A4} (1989) 2125;
F. David, Mod. Phys. Lett {\bf A5} (1990) 1019.}
\lref\greenseib{M. Green and N. Seiberg, Nucl. Phys. {\bf B299}
(1988)559;
M. Dine and N. Seiberg, Nucl. Phys. {\bf B301}(1988)357.}
\lref\goul{M. Goulian,``The Ising Model on a Fluctuating Disk,''
Santa Barbara preprint UCSBTH--91--01.}
\lref\cmnp{A. Cohen, G. Moore, P. Nelson, and J. Polchinski,
Nucl. Phys. {\bf B247}(1986)143;
A. Cohen, G. Moore, P. Nelson, and J. Polchinski,
``An Invariant String Propagator,'' in {\it Unified
String Theories}, M. Green and D. Gross, eds. World Scientific,
1986}
\lref\moore{G. Moore,``Double--Scaled Field Theory at $c=1$,''
Rutgers preprint RU--91--12.}
\lref\KM{D. Boulatov, V. Kazakov, I. Kostov and A. Migdal, Nucl. Phys.
B257 (1985) 433; F. David, Phys. Lett. 159B (1985) 303;
I. Kostov and M. Mehta, Phys. Lett. 189B (1987) 118.}
\lref\KDpl{B. Duplantier and I. Kostov, Nucl Phys.B340 (1990) 491.}
\lref\kostov{I. Kostov,``Strings embedded in Dynkin Diagrams,''
Saclay preprint SPhT/90--133, Lecture given at the Carg\`ese meeting.}
\lref\mehta{M. Mehta, {\it Random matrices}, Academic Press, 1991.}
\lref\mrcrg{G. Moore, Comm. Math. Phys. {\bf 133} (1990) 261;
``Matrix Models of 2D Gravity and Isomonodromic Deformation,'' to
appear in the proceedings of the 1990 Yukawa International Seminar, and
in the proceedings of the Carg\`ese workshop on Random Surfaces.}
\lref\gr{I.S. Gradshteyn and I.M. Ryzhik, {\it Table of Integrals, Series
and Products}, Academic Press, 1980.}
\lref\ab{M. Abramowitz and I. Stegun, {\it Handbook of Mathematical
Functions}, Dover, 1972.}
\lref\lizu{B. Lian and G. Zuckerman, ``New Selection Rules and Physical
States in 2d Gravity,'' Yale preprint YCTP-P18-90}
\lref\ctthrn{T.L. Curtright and C.B. Thorn, \prl {\bf 48} (1982) 1309;
E. Braaten, T. Curtright and C. Thorn, \pl {\bf 118B}
(1982) 115; Ann. Phys. {\bf 147} (1983) 365;
E. Braaten, T. Curtright, G. Ghandour and C. Thorn, \prl {\bf 51}
(1983) 19; Ann. Phys. {\bf 153} (1984) 147.}
\lref\gervais{J.-L. Gervais and A. Neveu, \np
{\bf 199} (1982) 59; {\bf B209} (1982) 125;{\bf B224} (1983) 329;
{\bf 238} (1984) 125; 396; \pl {\bf 151B} (1985) 271;
J.-L. Gervais, LPTENS 89/14; 90/4.}
\lref\jakiew{E. D'Hoker and R. Jackiw, \prl {\bf 50} (1983) 1719; \pr {\bf D26}
(1982) 3517; E. D'Hoker, D. Freedman and R. Jackiw, \pr {\bf D28}
(1983) 2583.}.
\lref\plsltw{A. Polyakov, Mod. Phys. Lett. A {\bf 2} (1987) 893.}
\lref\grklnw{D. Gross, I. Klebanov, and M. Newman, ``The two-point
correlation function of the one-dimensional matrix model,''
PUPT-1192}

\Title{RU-92-6}
{Multicritical
Phases of the O(n) Model
 on a Random Lattice }

\centerline{Ivan K. Kostov \footnote{$ ^\ast $}{on leave of absence
from the Institute for Nuclear Research and Nuclear Energy,
Boulevard Trakia 72, BG-1784 Sofia, Bulgaria}}

\centerline{Service de Physique Th\'eorique
\footnote{$ ^\dagger$}{Laboratoire de la Direction des Sciences
de la Mati\`ere du Comissariat \`a l'Energie Atomique} de Saclay
CE-Saclay, F-91191 Gif-Sur-Yvette, France}

\bigskip\centerline{Matthias Staudacher}

\centerline{Department of Physics and Astronomy
Rutgers University, Piscataway, NJ 08855-0849}

\vskip .3in

{\baselineskip10pt
We exhibit the multicritical phase structure of the loop gas model
on a random surface. The dense phase is reconsidered, with special
attention paid to the topological points $g=1/p$. This phase is
complementary to the dilute and higher multicritical phases in the
sense that dense models contain the same spectrum of bulk operators
(found in the continuum by Lian and Zuckerman) but a different set of
boundary operators. This difference illuminates
the well-known $(p,q)$ asymmetry of the matrix chain models.
Higher multicritical phases are constructed, generalizing both
Kazakov's multicritical models as well as the known dilute phase models.
They are quite likely related to multicritical polymer theories
recently considered independently by Saleur and Zamolodchikov.
Our results may be of help in defining such models on {\it flat}
honeycomb lattices; an unsolved problem in polymer theory.
The phase boundaries correspond again to ``topological'' points with
$g=p/1$ integer, which we study in some detail. Two qualitatively
different types of critical points are discovered for each such $g$.
For the special point $g=2$
we demonstrate that the dilute phase $O(-2)$ model does {\it not}
correspond to the Parisi-Sourlas model, a result likely to hold as well for
the flat case. Instead it is proven that the first {\it multicritical}
$O(-2)$ point possesses the Parisi-Sourlas supersymmetry.}

\vskip 1cm
\leftline{submitted for publication to {\it Nuclear Physics B}}
\rightline{SPhT/92-025}
\Date{3/92}

\newsec{Introduction and Overview}

\subsec{Introduction}

The past few years have seen considerable progress towards understanding
theories of conformal matter coupled to 2D gravity. Although some
insights were due to advances in continuum Liouville theory, stunning
progress was made in understanding and solving {\it discrete} models of
2D gravity. Initially, the number of such models available was small.
Due to recent interest in the subject, however, we now have a plethora
of such models available. Some of these models are known to correspond
to conformally coupled matter of central charges $C \leq 1$, while
the continuum limit of many others still remains to be understood.
Classifying the possible critical behaviour of lattice models of
2D gravity is potentially an important pursuit. Because of their
interpretation as toy models of bosonic string theory we might hope
to eventually attack more physical string theories with higher central
charges and built-in supersymmetry.

The perhaps simplest such theory is the one matrix model and its
multicritical points discovered by Kazakov
\ref\kaz{V.Kazakov, Mod.Phys.Lett A4 (1989) 2125.}.
They were identified
in \stau, \bdks as corresponding to the $(2m-1,2)$ minimal models coupled to
gravity. Even though this identification is beyond any doubt correct,
it is less clear why the lattice curvature dependent Boltzmann weights
of these models conspire to give precisely these minimal models. In
fact, we only understand this puzzle for the first three cases
$m=1,2$ and $m=3$ \stau. General $(p,q)$ minimal models require several
linearly coupled matrices \mike; actually two are sufficient
\ref\tmm{M.Douglas, ``The two-matrix model'', to appear in the
proceedings of the 1990 Carg\`ese workshop;
E.Martinec, Comm.Math.Phys. 138 (1991) 437;
T.Tada, Phys.Lett. B259 (1991) 442.}. Again
a better understanding of why the Boltzmann weights of these
multimatrix models result in particular conformally coupled theories
would be quite interesting.

An alternative way to introduce  $C \leq 1$  matter fields onto random lattices
is the loop gas construction \Ion \Iade. An advantage over the
multimatrix
approach is that the construction is manifestly physical; i.e. it is
clear (through the Coulomb gas mapping) why a specific critical behavior
occurs. All $(p,q)$  minimal models coupled to quantum gravity can be
constructed in this way.  So far a non-perturbative
definition of all these models  has not been given; it
 is however possible to derive diagrammatic
rules signaling the existence of a ``string field theory''
\ref\ivanlat{I.K Kostov, ``Strings with Discrete Target Space,'' Saclay
preprint SPhT/91-142.}.

The simplest statistical model allowing interpretation in terms of the loop
gas is the $O(n)$  model  \ref\Nienh{B. Nienhuis, in Phase Transitions and
Critical Phenomena, Vol. 11, C. C. Domb and J. L. Lebowitz, eds., Ch. 1
 (Academic Press, 1987)}. The $O(n)$ on a fluctuating surface  is
equivalent to a special one-matrix model  \Ion \ and thus can be defined
in a nonperturbative way.
 In this simplest model all loops
(contractible and noncontractible) are taken with the same weight $n$.
The partition function of the  corresponding
loop gas on a 3-coordinated fluctuating lattice
reads
\eqn\pflg{F(\lambda, T, N)= \sum _{\phi ^{3} {\rm graphs}} \ \
 \sum _{ {\rm loops}}
N^{2-2H} n^{\# {\rm loops}} e^{-\lambda v} e^{- T(
{\rm total\ length\ of\ loops})}}
where $\lambda$ is the cosmological constant coupled to the volume
$v = \#$ vertices  of the lattice,
 $T$ is the temperature of the loop gas
and $1/N$ is the string interaction constant coupled to the Euler
characteristics  $2-2H$ of the  graph.
The right hand side of  \pflg \
represents a triple series:
It is asymptotic in $1/N^2$ and of finite radius of convergence in
the fugacities
$g_{1}=e^{-\lambda}$ and ${1 \over 2P_{0}} = e^{-\lambda - T}$ corresponding
to vertices empty and occupied by loops, respectively.

For $N$ and $n$ integers this series coincides with the perturbative
expansion of the  vacuum energy of a zero-dimensional $N \times N$ matrix
field theory \Ion
\eqn\mm{Z={\int \cal D}M \prod_{i=1}^{n}{\cal D}\Phi_{i}
\exp \{ -N {\Tr} [V(M)+{1 \over 2} \sum_{\mu=1}^{n}\Phi_{\mu}^2-
{1 \over 2 P_{0}} \sum_{\mu=1}^{n} \Phi_{\mu}^2 M] \}}
The corresponding Feynman diagrams can be interpreted in terms of
surfaces populated by nonintersecting loops ( \fig\exlg{A
 configuration of nonintersecting loops })
Then  $n$ gives the number of species of loops and the
matrix potential  $V(M)$ specifies the measure in the  configuration space of
the random surface.
 With the choice $V(M)={1 \over 2} M^2
+ {1 \over 3} g_{1} M^3$ the perturbative expansion of the matrix integral
creates the cluster expansion of the $O(n)$ model on a 3-coordinated lattice.

The integration over the  matrix variables $\Phi _{\mu} , \mu =1,2,...,n$
can be performed
immediatelly and the result is the following one-matrix integral  \Ion\
\eqn\mim{Z={\int \cal D}M  \exp \ \{-N {\Tr} \ V(M)  +
{n \over 2} \int _{0}^{\infty}
{d \l  \over \l } ({\Tr} \ e^{M\l})^{2}\}}

The method of orthogonal polynomials used in the ``ordinary''
matrix models is not applicable here but
 the saddle point method in the limit $N \to \infty$ works equally well.
Indeed, the action depends only on the eigenvalues $P_{1},...,P_{N}$
of the random matrix $M$ and the integral \mim\  equals, up to a constant
factor, to
\eqn\mimd{Z=\int \prod _{i=1}^{N} dP_{i}  \ e^{-N\sum _{i=1}^{N} V(P_{i})
\ + \sum _{i \ne j} \log  |P_{i}-P_{j}| - {n \over 2}
 \sum _{i,j} \log | 2P_{0}-P_{i}-P_{j}| }}
 This last integral is defined   for any real value of $n$
and it is known from the flat case that criticality occurs for
$-2\leq n \leq 2$. This is exactly the interval in which the
saddle point solution exists \Ion . In this paper we will  restrict
our study to the
saddle point solution; this is sufficient for classifying the possible
critical regimes of the model.  The stability if the saddle point solution
and the nonperturbative effects due to instantons are considered in \ref\jzj{
B. Eynard and J. Zinn-Justin, in preparation}.

It is  very useful to parametrize $n$ in terms of the
Coulomb gas coupling constant $g$ as
\eqn\n{n=-2 \cos \pi g}
The model can be in a dense phase with $0 \leq g \leq 1$ or, if
$g_{1}$ in $V(M)$ is tuned appropriately, in a dilute phase with
$1 \leq g \leq 2$. In both phases the central charge is
given by $C=1 - 6 (\sqrt{g}- {1 \over \sqrt{g}})^2$.
Which models may be described within the two phases? Let us agree
that for our entire discussion in the present work
$p > q$. Then in the dense phase
$g={q \over p}$ and all minimal\foot{
Note that the $O(n)$ model is not minimal at generic values of $n$,
even if the associated central charge is in the BPZ list. Special
weights have to be assigned to noncontractible loops, as soon
as the genus exceeds zero or operator insertions are present
(see \Iade). In particular eq.\mm may not be used to give a
nonperturbative definition for the minimal models.}
models may be obtained; while
in the dilute phase $g={p \over q}$ and only models with
$p < 2q$ can be reached. In particular, the $p=2m-1$,$q=2$ models are
not present for $m \geq 3$. It is therefore natural to put them in
by hand and investigate the loop gas in the presence of a general
potential
\eqn\pot{V(M)=\sum_{k=2}^{m+1} {1 \over k} g_{k} M^{k}}
In order to analyze the possible critical behavior it is sufficient to
study the theory on a disc, i.e. we will consider the loop function
\eqn\res{W(P)=\langle {1 \over N} \Tr {1 \over P-M} \rangle}
where the average is taken with respect to \mm. For later use we
remind the reader that the asymptotics of
$W(P)$ for $P \rightarrow \infty$ is given by
\eqn\expnsn{W(P)={1 \over P}+{W_{1} \over P} +{W_2 \over P} + \ldots}

\subsec{Overview, Conclusions and Open Problems}

Our main result consists in establishing the phase diagram of
\fig\phases{The multicritical phase diagram of the loop gas coupled
to 2D gravity. $C$ is the central charge, $g$ is the Coulomb gas
coupling and $n$ the weight of the loops.}.
In all phases the relation $n=-2 \cos \pi g$ remains valid; i.e.
the phases may be thought of as an infinite number of branches
in the complex $n$-plane (the branchpoints being at $n=\pm 2$).
The Kazakov multicritical points are situated by construction
at the center (i.e. $n=0$) of each phase. For a given central
charge between $-\infty$ and $1$ there are always two points; one
lying in the dense and the other in a higher phase. These two models
related by the duality $g \rightarrow {1 \over g}$ are however
{\it not} equivalent. This is particularly surprising for the
minimal models: There is a dense realization at $g={q \over p}$
and a multicritical one at $g={p \over q}$. We argue in section 2.2.
that the difference is due to a different set of {\it boundary}
operators. From this point of view, then, the infamous $(p,q)$
asymmetry of the matrix chain models arises because the chain models
always favor one set of boundary operators over the other
(with the exception of the Hermitian one matrix model with
$g={1 \over 2}$).
Another sequence of particularly interesting points in the dense
phase is generated by the ``topological'' series $g={1 \over p}$.
We demonstrate that these models appear in our lattice approach as
theories of ``boundaries'' (i.e. loops) without ``worldsheet''
\foot{This way of looking at topological theories was first clearly
stated in \ivanlat,\mss.}.
Our understanding is aided by the realization that the loop equation
of the model may actually be {\it exactly} solved for rational
values of $g$.
The first member ($g={1 \over 2}$)
of this series is the model of topological gravity whose continuum
limit is known \ref\dist{J.Distler, Nucl. Phys. B342 (1990) 523.}.
The correct continuum limit of the higher
members of the series remains to be found.
In section 3. we further establish the above diagram
by investigating the
``perturbation'' of the $(2m-1,2)$ models by the loop gas. Note that our
approach smoothly interpolates between the physical construction
in the dense and dilute phase and the higher Kazakov points. One may
therefore hope to eventually obtain a better understanding of the
latter. Let us note that flat multicritical $O(n)$ models have recently
attracted some attention in relation to a search for multicritical
theories of polymers \ref\salpol{H.Saleur, ``Polymers and Percolation
in Two Dimensions and Twisted $N=2$ Supersymmetry,'' Yale preprint
YCTP-P38-91;
H.Saleur, ``Geometrical Lattice Models for $N=2$ Supersymmetric
Theories in Two Dimensions,'' Yale preprint YCTP-P39-91.},
\ref\zam{A.B. Zamolodchikov, unpublished.}.
Polymers are obtained by taking a
$n \rightarrow 0$ limit in the $O(n)$ model. Dense and dilute polymers
have been related to a, respectively, broken and unbroken twisted
$N=2$ supersymmetry \salpol, and it is natural to ask whether some insights
into 2D gravity and noncritical string theory can be derived from this fact.
A further very interesting problem is to find out what kind of multicritical
polymer theories are
obtained by taking
an $n \rightarrow 0$ limit in the Kazakov backgrounds. A comparison
of critical exponents (see 3.3.) indicates that our multicritical
polymers do not correspond to the ones considered in \salpol.
In section 4. the phase boundaries between the various regimes are
studied. A discontinuity at these points is detected,
resulting in two qualitatively different models at each integer $g$.
The first boundary, $g=1$, possesses $C=1$, while the second
boundary, $g=2$, has $C=-2$. We demonstrate that the model
connected to the $m=3$ phase possesses the Parisi-Sourlas
``target space'' supersymmetry. The dilute $O(-2)$ model is different;
only a $n \rightarrow -2$ {\it limit} in the dilute phase yields
again the Parisi-Sourlas model. The further boundaries at
$g=3,4,\ldots$ all correspond to ``topological'' models. Their correct
continuum description remains obscure for the moment.

\newsec{The Dense Phase, Revisited}

\subsec{The Loop Equation and its Solution}

The dense phase corresponds to setting all couplings $\{g_k\}$ in
\pot\ to zero, except for $g_0=1$. All loops are then densely packed
on the $\varphi^3$ random graphs. The loop equation reads
\eqn\lsmpl{W^{2}(P) = \oint _{\cal C} {dP' W(P') \over 2\pi i (P-P')}
[V'(P')-nW(2P_{0}-P')]}
It may be derived from the matrix model \mm or by combinatorial
methods \Iade. After a symmetrization with respect to the reflection
$P \to 2P_{0}-P$
the contour integral can be performed using the Cauchy theorem  and
we arrive at the following functional equation ( for the
gaussian potential $V(P)={1 \over 2} P^{2}$ )
\eqn\leqdns{
W(P)^2 + W(2 P_{0} - P)^2 + n W(P) W(2 P_{0}-P)=
P W(P) + (2 P_{0}-P) W(2 P_{0}-P) - 2
}
 The critical point is located to be at
$P_{*}=\sqrt{2(2+n)}$. If we transform $P=P_{0}+{\bar P}$ and
$W(P)={1 \over 2 \sin^2 \pi g}[P+\cos \pi g\;\;
(2 P_{0}-P)]+{\bar W}({\bar P})$ \leqdns simplifies to
\eqn\leqsimp{{\bar W}({\bar P})^2+{\bar W}(-{\bar P})^2
+n {\bar W}({\bar P}){\bar W}(-{\bar P})=
{1 \over 2+n}(P_{0}^2-P_{*}^2)+{1 \over 2-n}{\bar P}^2}
In \Icar ,\Iml\  the equation was solved in the
vicinity of $P_{*}$. After introducing a cutoff ${\bar a}$  we blow up the
vicinity  of  $P_{\ast}$ by the following change of variables
\eqn\sclim{{P_{0}-P_{*} \over P_{\ast}}=   {\bar a}^{2g} \Lambda ,\
\ {P-P_{0} \over P_{\ast}-P_{L}} = {\bar a} z, \ \ {P_{0}-P_{R}
\over P_{\ast}-P_{L}} = {\bar a } M }
Here $\Lambda$ is the continuum cosmological constant,
$z$ the continuum boundary cosmological constant, and $M$ is the
boundary cosmological constant induced by the fluctuations of the
worldsheet geometry.

One obtains for
the singular part ${\bar a}^g w(z)$ of $W(P)$
\eqn\wzdns{\eqalign{
w(z)&= -{1 \over 2} A_{g}
 [ (z + \sqrt{z^2 - M^2})^g +
(z - \sqrt{z^2 - M^2})^g ] \cr
&= A_{g} \  M^{g}\  \cosh (g\tau);\cr
 \ \ \ \ \  z&=M\ \cosh (\tau) \cr }}
with
\eqn\mlmbda{\Lambda= {2 \over (1-g)^2}
M^{2g}}
\eqn\klklkl{A_{g}={- 2 \sqrt{2} \over 1-g} \ \  {1  \over \sin \pi g}}
The last relation follows if we compare \wzdns\ with the exact solution
obtained for $P_{0}=P_{\ast}$ in  \ref\gauivan{M.Gaudin and I.Kostov,
Phys.Lett. B220(1989)200.}
\eqn\crsln{\eqalign{
\bar {W} &=- {1 \over  2} A_{g}\  {u^{1-g}+u^{g-1}
\over u + u^{-1}} \cr
\bar {P} & = {2(P_{\ast}-P_{L}) \over u+u^{-1}} \cr
P_{\ast} -P_{L}  &={\sqrt{2(2-\beta )}  \over 1-g} \cr
}}

The inverse Laplace image of $w(z)$ gives the amplitude $\tilde w(\ell )$
for a disc with fixed length $\ell $ of its boundary
$$w(z)= \int _{0}^{\infty} d\ell e^{-z\ell} \tilde w(\ell);$$
\eqn\qwer{
\tilde w(l)={2^{3/2} g  M^{g} \over  (1-g)} {1 \over \ell } K_{g}(\ell)
\sim
\cases{
{2g \sqrt{\pi} \over (1-g)}
M^{g-1/2} \ell ^{-3/2} e^{-M\ell},  &if $\ell \gg 1/M, $ \cr
-{2^{g+1/2} \pi  g \over (1-g) \sin \pi g}  M^{-g} \ell ^{-g-1},
   &if $\ell \ll 1/M, $ \cr}}

Now \wzdns, \mlmbda\ is quite puzzling for several reasons:

1. \mlmbda\  says that quantum area {\it does not} scale like the
square of quantum length in these models. An attempt at an
explanation is made in section 2.2.

2. For the topological models $g={1 \over p}$ we see that the
string susceptibility is an integer: $\gamma_{s}=1-{1 \over g}$.
This is surprising, since it means {\it either} that the exact
lattice solution is analytic in the lattice cosmological
constant $P_{0}$ around $P_{*}$ {\it or} that we have
logarithmic scaling violations at these points. Below we
will argue that the first possibility holds (in accordance
with \ref\sal{H.Saleur, Mod.Phys.Lett. A6 (1991) 1709.}).

3. A subtle but important point is that, for $g< {1 \over 2}$,
we have to approach the critical
value of the boundary cosmological constant as
$P=P_{0}+{\bar a} z$, not $P=P_{*}+{\bar a} z$. We do not have
a deep explanation for this, but note that it constitutes
annother instance of an ``analytical redefinition'' (discovered
in \mss) necessary to get sensible results from lattice
gravity models.

It is interesting to realize that \leqdns may actually be exactly
solved for {\it rational} $g$.
This serves as a check for \wzdns, \mlmbda and will help
in getting some insight into the second problem just mentioned.
At the critical point $P_{0}=P_{*}$ the solution to \leqsimp has been
known already for some time  \gauivan
 \Iade. It may be parametrized using
circular functions or rational functions($g={q \over p}$):
\eqn\wpgaud{\eqalign{&{\bar W}=-{\sqrt{2} \over 1-g} \ {1 \over \sin \pi g}
\;\;{t^{2p-q}+t^q \over t^{2p}+1} \cr
&{\bar P}={4 \over 1-g}\sqrt{1+\cos \pi g}\;\;{t^p \over t^{2p}+1} \cr}}
Here $t$ is defined on (the double cover of) the Riemann sphere. It is
apparent from \wpgaud that the Riemann surface of $W(P)$ is
{\it algebraic}. Furthermore, since the surface is uniformized by
rational functions of $t$ we see that it has genus zero.
$P$ is defined on a surface with $p$ sheets. On the lowest (physical)
sheet there is one cut from $P_{L}$ (branchpoint of order 2)
to $P_{R}=P_{0}$ (branchpoint of order p). The higher (unphysical)
sheets possess also a cut from $P_{0}$ to $2 P_{0}-P_{L}$, except for
the topmost, which again has just one cut. Now it is important
to understand what happens if we move away from the critical point:
The degeneracy at $P=P_{0}$ will be removed and all branchpoints
will be of order 2. However, we still have a Riemann surface; it has
a finite number of sheets and only square root singularities.
By a standard theorem we therefore conclude that the solution of \leqsimp
is algebraic and thus given by
\eqn\solwp{{\bar W}^p+h_{p-1}({\bar P}) {\bar W}^{p-1}
+\ldots+h_{1}({\bar P}) {\bar W}+h_{0}({\bar P})=0}
Here the $h_{i}({\bar P})$ are {\it polynomials} in ${\bar P}$, with $P_{0}$
dependent coefficients. These coefficients are determined
by expanding \leqsimp, \solwp around ${\bar P}=\infty$ and matching
the two expansions. It is easy to see that the genus of the
Riemann surface will not change off the critical point; we
therefore know the existence of a parametrization of \solwp
using rational functions of one parameter. Unfortunately we did not
succeed in finding a simple such parametrization for arbirary $g$.
Nevertheless we are now in a position to get some insight into
the second problem mentioned above. Given the just presented algorithm for
generating exact solutions to the loop equation we see that
the singularities in the lattice cosmological constant $P_{0}$
are necessarily {\it algebraic} for rational $g$. Logarithmic
scaling violations are thus impossible and we have {\it analytic}
behavior in $P_{0}$ at the topological points $g={1 \over p}$.

\subsec{Bulk and Boundary Operators}

The loops in 2D quantum gravity may be considered as generating
functions for the local operators of the theory\bdss. Since
the loops of the dense phase are clearly different from the ones
in the standard lattice models it will be interesting to investigate
their spectrum. Recall that Lian and Zuckerman have calculated the
bulk spectrum in a version of continuum Liouville gravity. They found
for the $(p,q)$ minimal model coupled to gravity an infinite number
of states with Liouville charges $\alpha$ given by (as always $p>q$)
\eqn\lz{{\alpha \over \gamma}={p+q-k \over 2 q}\;\;\;\;\;k=1,2,\ldots}
with the important restrictions $k\not\equiv 0(\rm{mod}\;q)$ and
$k \not\equiv 0(\rm{mod}\;p)$ ($\gamma$ is the charge of the identity
operator, $k=p-q$). In \ref\kusei{D.Kutasov and N.Seiberg,
Nucl.Phys. B358 (1991) 600},\bk
it was shown that it is precisely these states which
propagate around the
torus. Now, in \ivanlat it was found that the
torus partition function for the dense minimal models coincides with
the one found in \bk, establishing that the dense models do have the
expected bulk spectrum. As is well known, the standard matrix model
realizations of the minimal models contains further operators. One way
to see them is to look at the scaling dimensions obtained from
expanding the macroscopic loops in the length $\ell$. It is found that
there are also operators with charges \lz but $k$ divisible by $p$. They were
identified by Martinec, Moore and Seiberg \ref\emil{
E.Martinec, G.Moore and N.Seiberg, Phys.Lett. B263 (1991) 190.}
as boundary operators. The first such operator, $k=p$, has
$\alpha={\gamma \over 2}$ and is beautifully interpreted as the length
($\phi$ is the Liouville field)
\eqn\lplu{\ell= \oint e^{{\gamma \over 2} \phi}}
For the dense phase models, the spectrum of dimensions found in the
loop expansions was worked out in \Iml.
It is easily seen that they correspond to
operators with ${\alpha \over \gamma}$ as in \lz but with $k$
divisible by $p$ {\it missing} and $k$ divisible by $q$
{\it appearing}. The dense models are therefore different from the
standard models because they possess a different set of boundary
operators. This has dramatic effects, like the non-trivial
Hausdorff dimension of the boundary. Indeed, here the first
boundary operator has $\alpha={p \over q}{\gamma \over 2}$ and may
be interpreted as the quantum length
\eqn\lmin{\ell= \oint e^{{p \over q} {\gamma \over 2} \phi}}
The Hausdorff dimension $2g=2{q \over p}$ may be directly read off
from the last equation. Note that the choice of $\alpha$ in
\lmin corresponds to taking the ``wrong'' branch in the KPZ
dressing of the identity on the boundary. It appears that the
Seiberg rule for choosing these branches does not necessarily
hold for boundary operators \nati \emil.
It should also be pointed out that the dense minimal models possess
$p-1$ loops as opposed to the dilute minimal models, which have
$q-1$ loops. In the KP description of the minimal models inspired
by the matrix chains the dense phase is formally recovered by
interchanging the operators $P$ and $Q$ (see \mike, \tmm,\kutdi).
This remains however a
formal excercise since it is not hard to see\foot{
We acknowledge a discussion with M.R.Douglas on this matter.}
that for the matrix
chains the order of $P$ is always larger than the order of $Q$ (except
for the hermitian one matrix model which is in the dense phase.).
Our interpretation of
dense models as being endowed with ``standard worldsheets'' but
``nonstandard boundaries'' is especially manifest in the example
of the subsequent section.

\subsec{$g={2 \over 3}$}

$g={2 \over 3}$ is ``pure gravity'' ($C=0$) in the dense phase. And
indeed, for $n=1$ in the maximally dense case (only $g_{0}=1\neq 0$)
we can integrate out the $M$ matrix in \mm and obtain
\eqn\mmf{Z=\int {\cal D}\Phi_{1}
\exp{\{ -N {\Tr} [{1 \over 2}\Phi_{1}^{2} - {1 \over 4 P_{0}^2}
\Phi_{1}^{4}] \} } }
Of course the integration is this trivial only for the
partition function and gets harder (although it may still be done) if
the boundary is present: Remember the loop operator is
${1 \over N}{\Tr} e^{LM}$, {\it not} ${1 \over N} {\Tr} e^{L\Phi_{1}}$.
It shows, however, that the {\it bulk} consists of ordinary
$\varphi^4$ lattice gravity while the {\it boundary} is different.

The exact solution \solwp to the loop equation \leqsimp reads
\eqn\alzwdrei{{\bar W}^3-({1 \over 3}(P_{0}^2-6)+{\bar P}^2 )
{\bar W} - {2 \over 27}(P_{0}^2-6)^{3 \over 2}+{2 \over 3}P_{0}
{\bar P}^2=0}
Being of third degree, it may be solved for ${\bar W}({\bar P})$ to yield
($\Delta=P_{0}^2-6$)
\eqn\solzwdrei{\eqalign{&{\bar W}({\bar P}) = \cr
&-\Biggl[{1 \over 3} P_{0} {\bar P}^2 - {1 \over 27} \Delta^{3 \over 2} +
{1 \over \sqrt{27}} {\bar P} \sqrt{2 (P_{0}^2+3) {\bar P}^2 -
{\bar P}^4 - {1 \over 3}\Delta^2- {2 \over 3}P_{0}\Delta^{3 \over 2}}
\;\;\Biggr]^{1 \over 3} \cr
&-\Biggl[{1 \over 3} P_{0} {\bar P}^2 - {1 \over 27} \Delta^{3 \over 2} -
{1 \over \sqrt{27}} {\bar P} \sqrt{2 (P_{0}^2+3) {\bar P}^2 -
{\bar P}^4 - {1 \over 3}\Delta^2- {2 \over 3}P_{0}\Delta^{3 \over 2}}
\;\;\Biggr]^{1 \over 3} \cr}}

\subsec{$g={1 \over 3}$}

This model is the first nontrivial point in the ``topological''
series $g={1 \over p}$ and has $n=-1$. Because of the minus signs
one easily proves that all {\it closed} string diagrams are identically
zero because of precise cancellations. The free energy is zero even
before taking the scaling limit! Open string diagrams, however,
survive (see \fig\topdr{An example for a diagram of the topological
model $(1,3)$.}). Just as $g={1 \over 2}$, the model
is rather a theory of loops than of surfaces. The solution \solwp of
\leqsimp is
\eqn\aleidrei{{\bar W}^3-(P_{0}^2-2 +{1 \over 3} {\bar P}^2) {\bar W}
+{2 \over 3}(1+P_{0}^2) {\bar P} - {2 \over 27} {\bar P}^3=0}
The equation may be solved to give ($\Delta=P_{0}^2-2$)
\eqn\soleidrei{\eqalign{&{\bar W}({\bar P})= \cr
&+\Biggr[{1 \over 27}{\bar P}^3- {1 \over 3}(P_{0}^2+1){\bar P}+
\sqrt{{1 \over 9}(P_{0}^2+1)^2 {\bar P}^2-{1 \over 27}\Delta^2{\bar P}^2
- {1 \over 27}P_{0}^2 {\bar P}^4- {1 \over 27}\Delta^3 }
\Biggl]^{1 \over 3} \cr
&+\Biggr[{1 \over 27}{\bar P}^3- {1 \over 3}(P_{0}^2+1){\bar P}-
\sqrt{{1 \over 9}(P_{0}^2+1)^2 {\bar P}^2-{1 \over 27}\Delta^2{\bar P}^2
- {1 \over 27}P_{0}^2 {\bar P}^4- {1 \over 27}\Delta^3 }
\Biggl]^{1 \over 3} \cr }}
It is manifest that the solution is analytic at $P_{0}^2=P_{*}^2=2$.

\subsec{$g={1 \over 4}$}

In order to study the puzzling question of the singularity in the
lattice cosmological constant we have analyzed yet annother topological
point. The model is less trivial than $g={1 \over 3}$ because
here $n=-\sqrt{2}$ and
we do not have precise cancellation of the worldsheet-loops away from
the scaling region. The exact solution for ${\bar W}({\bar P})$ is
(we introduced the abbreviations $\sigma_{\pm}=1 \pm {1 \over 2} \sqrt{2}$)
\eqn\soleivier{{\bar W}({\bar P})=
-\sqrt{\sigma_{+} (P_{0}^2- 4 \sigma_{-}) + \sigma_{-} {\bar P}^2
-{1 \over 2} \sqrt{2} \sigma_{-} ({\bar P}- B)
\sqrt{({\bar P}-{\bar P}_{R})({\bar P}-{\bar P}_{L})} } }
which is, in accordance with \solwp a solution to a bicubic equation for
${\bar W}$. The positions of the cut are given through
${\bar P}_{R} + {\bar P}_{L}= 4 \sqrt{2} \sigma_{+} P_{0} - 2 B$ and
${\bar P}_{R} {\bar P}_{L}=4 \sigma_{+}^4 (P_{0}^2-4 \sigma_{-})^2
{1 \over B^2}$. The auxiliary parameter $B$ is determined by
\eqn\B{B^4- {8 \over 3}\sqrt{2} \sigma_{+} P_{0} B^3
+{4 \over 3} \sigma_{+}^2 (P_{0}^2 - 4 \sigma_{-}) B^2
-{4 \over 3} \sigma_{+}^4 (P_{0}^2 - 4 \sigma_{-})^2=0}
It is seen\foot{That is because the $j$-leg functions $\{W_j\}$
are simple rational functions in $B$, as \soleivier shows. E.g., $W_1$ is
given by $P_0 + {1 \over B^2}[{1 \over 4} \sqrt{2} B^5 -
(1 + {1 \over 8} \sqrt{2})P_0 B^4 - 2 \sigma_{+} P_{0}^2 B^3+
(2 \sigma_{+}^4 (P_{0}^2 - 4 \sigma_{-})P_0 - ({25 \over 2}+
{35 \over 4} \sqrt{2}) P_{0}^3)B^2 - \sqrt{2} \sigma_{+}^4 (P_{0}^2-4
\sigma_{-})^2 B - ({3 \over 4} \sqrt{2}+1) \sigma_{+}^2 (P_{0}^2 - 4
\sigma_{-})^2 P_0]$}
that any singular behavior in the cosmological constant
$P_{0}$ would have to be generated from eq.\B. A (most conveniently
numerical) study shows however that the branchpoints of the
{\it physical} sheet of \B are {\it not} located at the critical
value $P_{*}^2=4 \sigma_{-}$ (i.e.$P_{0}\approx 1.082$)
but are lying off the real axis at
$\pm0.327 \pm0.633 i$. This confirms the general discussion given in
section 2.1. and agrees with the picture of \sal. Let us further note
that, since there are no singularities on the real $P_{0}$ axis,
the lattice model does not undergo {\it any} phase transition in
the bulk cosmological constant, i.e. not even for some $P_{0} <
P_{*}$.

\newsec{The Multicritical Phases}

\subsec{Solution and Scaling of the Loop Equation}

We will now turn on $m-1$ couplings $\{g_k\}$ in order to reach the $m^{th}$
multicritical phase. The loop equation is easily generalized to
\eqn\leqmult{\eqalign{&W(P)^2+W(2 P_0 -P)^2+nW(P)W(2 P_0-P)= \cr
&=V'(P)W(P)+V'(2 P_0 -P)W(2 P_0 -P)+\sum_{k=1}^{m}
\sum_{j=1}^{k}g_{k+1}W_{j-1}[P^{k-j}+(2 P_0 - P)^{k-j}] \cr }}
where $V'$ is the derivative of the potential \pot. A new feature
compared to the dense phase equation \leqdns is the appearance of the
(as yet) unknown functions $\{W_j\}$ (see eq.\expnsn) which depend on all
matter couplings $\{g_k\}$,$P_0$. As for the dense phase the
linear terms may be eliminated through the transformation
$P=P_0+{\bar P}$, $W(P)={1 \over 2 \sin^2 \pi g}[V'(P)+
\cos \pi g\; V'(2 P_0 - P)] + {\bar W}({\bar P})$:
\eqn\leqmusi{{\bar W}({\bar P})^2+{\bar W}(-{\bar P})^2
+ n{\bar W}({\bar P}){\bar W}(-{\bar P})=
\sum_{i=0}^{m} s_{2i} {\bar P}^{2i} }
Here the $\{s_{2i}\}$ are both explicitly and implicitly (through the
functions $\{W_j\}$) dependent on the matter couplings. In the
$m^{th}$ critical phase the loop function should behave for
${\bar P} \rightarrow 0$ as
$W({\bar P}) \sim {\bar P}^g$ with $m-1\leq g \leq m$. The critical
point is found to be fixed through the conditions\foot{
Note that we have $m$ couplings at our disposal in order to
fulfill the $m$ conditions.}
\eqn\comult {s_{0}=s_2=\ldots=s_{2(m-1)}=0\;\;\;\;\;  s_{2m}\neq 0}
At that point eq.\leqmusi may be explicitly solved; the solution with
the correct asymptotics at ${\bar P}=\infty$ reads
\eqn\wpgamu{\eqalign{
{\bar W}({\bar P})&=-{1 \over 2 \sin \pi g}
{1 \over \sqrt{2-n}} g^{*}_{m-1} {\bar P}^m\;
\Bigg[\Bigg({-{{\bar P_{L}} \over {\bar P}}}  + \sqrt{
\big({{\bar P_{L}} \over {\bar P}}\big)^{2} -1} \Bigg)^{m-g} \cr
&+\Bigg({-{{\bar P_{L}} \over {\bar P}}}  - \sqrt{
\big({{\bar P_{L}} \over {\bar P}}\big)^{2} -1} \Bigg)^{m-g} \Bigg]
\cr}}
The position of the left branchpoint ${\bar P}_{L}$ as well as the
critical couplings $\{g^{*}_k\}$ are  determined from the asymptotics.
We see that indeed $W({\bar P}) \sim {\bar P}^g$. In view of this
solution at the $m$-th multicritical point our arguments about the
structure of the Riemann surface of ${\bar W}$ in section 2.1.
may be repeated;  we thus conclude that the exact solution for
{\it rational} $g={p \over q}$ and arbitrary matter couplings is given by
\eqn\solwpmu{{\bar W}^q + h_{q-1}({\bar P}) W^{q-1} + \ldots +
h_{1}({\bar P}) {\bar W} + h_{0}({\bar P})=0 }
Again the $h_i({\bar P})$ are polynomials in ${\bar P}$ whose degree
increases with $p$; they (as well as the functions $W_j$) are determined
by matching \solwpmu and \leqmusi at ${\bar P}=\infty$. In order to
infer the scaling limit of ${\bar W}({\bar P})$ we however do not need
this exact solution; instead we simply generalize
the arguments of \Icar, \Iml: upon scaling ${\bar P}={\bar a}z$ one
concludes from \wpgamu that ${\bar W}({\bar P}) \sim {\bar a}^g w(z)$;
thus the dominant piece in \leqmusi is $s_{0}$ which has to vanish
as ${\bar a}^{2g}$. Parametrizing the proper approach to the critical
point \comult by the continuum cosmological constant $\Lambda$ one
expects to find $s_{0} \sim {\bar a}^{2g} \Lambda^g$; this will give
as expected $w(z)={-1 \over \sin \pi g} [ (z + \sqrt{z^2 - \Lambda})^g +
(z - \sqrt{z^2 - \Lambda})^g ]$. The Hausdorff dimension takes on the
``classical'' value $2$ for all multicritical phases.

\subsec{Boltzmann Weights on Flat and Random lattices}

The dense and dilute loop gas has been constructed as a direct
adaptation of flat lattice models to the random case. It is therefore
natural to ask whether we can go back from our formulation of the
multicritical loop gas and deduce how a multicritical $O(n)$ model
might look like on {\it flat } honeycomb lattices. Generically we
would expect to obtain such a model by introducing interactions
between the loops \zam, \salpol. In the case of the first multicritical phase
$(m=3)$ a concrete suggestion may be made, using the procedure of \stau.
In this case the matrix model \mm generating the loop gas may be
rewritten with the help of an additional matrix $\Psi$ as
\eqn\mmd{\eqalign{&Z={\int \cal D}M {\cal D}\Psi
\prod_{i=1}^{n}{\cal D}\Phi_{i} \times \cr
& \times
\exp \{ -N \Tr [{1 \over 2} M^2 + {g_{1} \over 3} M^3 + {1 \over 2}
\Psi^2 + i \sqrt{{g_{2} \over 2}} \Psi M^2 +
{1 \over 2} \sum_{\mu=1}^{n}\Phi_{\mu}^2-
{1 \over 2 P_{0}} \sum_{\mu=1}^{n} \Phi_{\mu}^2 M] \} \cr }}
which is seen to possess the graphical expansion exemplified in
\fig\loopdim{A typical diagram in the $m=3$ multicritical phase.}.
In addition to the loops we have dimers placed on the random
``honeycomb'' lattice. Their fugacity has to be chosen negative
in order to perturb away from the $m=2$ regime. Note that this
formulation may be applied to flat lattices and, even though it does
not constitute an exactly solvable model, could be checked by numerical
analysis of the transfer matrix. According to our phase diagram in
\phases \
we predict central charges between $C=-2$ and $C=-7$. For the
higher phases it is less clear how to proceed. One guess would be to
complement the dimers by increasingly longer strands
(2-chains, 3-chains etc.) with alternating signs for their Boltzmann
weights. An interesting question from the point of view of polymer
theory is whether the multicritical matter induces in the continuum
limit an effective interaction between the loops. An alternative
possibility
is that directly introducing such interactions results in yet another
class of multicritical polymers.

\subsec{Geometrical critical exponents for the $O(n)$ vector model}

The order parameters (the magnetic operators)
 in the $O(n)$ model have a simple discription in terms of the loop gas.
 The $m$-th magnetic operator  $\Psi_{L}$  is represented
as the source of $L$  nonintersecting lines meeting at a point.
The correlation function of two such operators can be evaluated as the
partition function   $F_{L}$
 of a network consisting of $L$ nonintersecting lines
tied at their extremities, moving in the sea of vacuum loops of the $O(n)$
model \KDpl.
The dimension of the star operator $\Psi _{L}$ can be extracted from the
dimension of the partition function $F_L$. This last quantity
can be calculated immediately.

First observe that the $L$ nonintersecting lines cut the world sheet
into k pieces with the topology of a disk.  Let $\ell _{1},...,\ell _{L}$
be the lengths of the $L$ lines which form the watermelon network. As
usual, we first sum over all configurations of the world sheet populated
by vacuum loops keeping these lengths fixed.
Then the partition function $F_{L}$  can be represented as an integral
over $\ell _{1},... \ell _{L}$ of the product of $L$ loop amplitudes
with lengths $\ell _{1}+\ell _{2},\ell _{2}+\ell_{3},...,\ell _{L}+\ell _{1}$

$$ F_{L} = \int  \prod _{i=1}^{k} d\ell _{i}  e^{-2P_{0} \ell _{i}}
\ \ \  \tilde  w(\ell _{1}+\ell _{2})
 \tilde w(\ell _{2}+\ell _{3}) ... \tilde w(\ell _{L}+\ell _{1})    $$

By the general scaling argument

$$ F_{L} \sim  \Lambda ^{2\delta _{L} - \gamma _{str}} $$

where $\delta _{L}$ is the gravitational
dimension of the operator $\Psi _{L}$.
On the other hand  the loop amplitude behaves asymptotically as
  $w(\ell ) \sim \ell ^{-1-g}$ and therefore

$$ F_{L} \sim  M^{Lg} = \Lambda ^{Lg/(2\nu )} $$

It follows that

$$\delta _{L} = (Lg/(2\nu) + \gamma _{str})/2 =
\cases{
{Lg \over 4} - {g-1 \over 2}, & if $g>1$ \cr
{L \over 4} - {1-g \over 2g}, &if $g<1$  \cr} $$

The corresponding flat conformal dimensions are

$$ \Delta _{L}=
\cases {
 \Delta _{L/2,0} =
{g \ L^{2} \over 16} - {(g-1)^{2} \over 4g} , & if  $g>1$ \cr
\Delta _{0,L/2} ={L^{2} \over 16 g} - {(g-1)^{2} \over 4g}, &if
$g<1$ \cr} $$

The smallest dimension $\Delta _{1}$ is positive only in the interval
$-1/2 \le g \le 2$. Outside this interval the propagator of the
nonintersecting random walk grows with the distance between its two
extremities. The fact that two points are connected with a line
leads to an effective repulsion between them. Such a phenomenon is
typical for nonunitary theories.

\newsec{The Phase Boundaries}

\subsec{Exact Solution and Scaling of the Loop Equation at $n=\pm2$}

We will now investigate in some detail the loop gas with $n=\pm2$
in the presence of multicritical matter. This will serve as an
important check of our ideas. As we discussed before we expect
to be able to tune to points with $g=p$ integer, which form the
crossover points from the $p$-phase to the $(p+1)$-phase.
Consider again the scaled loop function
\eqn\wz{w(z)={-1 \over \sin \pi g} [ (z + \sqrt{z^2 - M^2})^g +
(z - \sqrt{z^2 - M^2})^g ] }
Were it not for the ``wavefunction renormalization'' factor
${1 \over \sin \pi g}$ we would get a trivial (i.e. purely analytic)
result for $w(z)$ for $g \in N$. Including this factor and subtracting terms
analytic in $z$ we instead
obtain
\eqn\wzn{w(z)={(-1)^{g-1} \over \pi} [ (z+\sqrt{z^2-M^2})^g -
(z-\sqrt{z^2-M^2})^g ] \log{{z+\sqrt{z^2-M^2} \over M}}}
Upon Laplace-transforming this result, one obtains correctly
$w(\ell)={1 \over \ell} M^g K_{g}(M \ell)$, as one expects from
Liouville theory \mss. We will now demonstrate how to derive this
result from our lattice model. The algebraic approach clearly breaks
down for integer $g$: the Riemann surface associated with $W(P)$
becomes ininitely foliated. Fortunately it is however possible
to directly solve the saddlepoint equation of the $O(n)$ model's
matrix model formulation \ref\gau{M.Gaudin,unpublished.}.
This equation for the eigenvalue density of the matrix model reads
\gauivan
\eqn\sp{\int_{a}^{b}dy \rho(y) ({1 \over x-y} + {n \over 2}
{1 \over 2 P_{0}-x-y})={1 \over 2} V'(x)=
{1 \over 2} \sum_{k=1}^{m} g_{k+1} x^k}
where $V(x)$ is the potential of the matrix model.
It is convenient to change variables to $\lambda'={1 \over 2}(P_{0}-x)$,
$\mu'={1 \over 2}(P_{0}-y)$, $b_{0}={1 \over 2} P_{0}$,
$\rho(x) \rightarrow {1 \over 2} \rho(\lambda')$, $g_{k} \rightarrow
(-{1 \over 2})^{k-2} g_k$ and rewrite \sp  as
\eqn\spz{\eqalign{&\int_{a'}^{b'} \rho (\mu') {2 \mu' \over \lambda'^2-
\mu'^2}=2 \sum_{k=1}^{m} g_{k+1}(\lambda' - b_{0})^k \;\;\;\;\; (n=+2) \cr
&\int_{a'}^{b'} \rho (\mu') {2 \lambda' \over \lambda'^2-
\mu'^2}=2 \sum_{k=1}^{m} g_{k+1}(\lambda' - b_{0})^k \;\;\;\;\; (n=-2)
\cr}}
A further transformation $\lambda'=\sqrt{A+B \lambda}$ with
$A={1 \over 2} (a'^2 + b'^2)$ and $B={1 \over 2} (a'^2 - b'^2)$
simplifies these equations to
\eqn\spd{\eqalign{&\int_{-1}^{1} d\mu {\rho(\mu) \over \lambda - \mu}
=2 \sum_{k=1}^m g_{k+1} (\sqrt{A+B \lambda} - b_{0})^k \;\;\;\;\;(n=+2)
\cr &\int_{-1}^{1} d\mu {\rho(\mu) \over \sqrt{A+B \mu}}
{1 \over \lambda - \mu}
=2 \sum_{k=1}^m {g_{k+1} \over \sqrt{A+B \lambda}}
(\sqrt{A+B \lambda} - b_{0})^k \;\;\;\;\;(n=-2) \cr}}
These integral equations may now be immediately inverted
\ref\musch{N.Muskhelishvili, Singular Integral Equations (Nordoff NV,
Groningen, Netherlands, 1953).} to yield
\eqn\sol{\eqalign{&\rho(\lambda)=-{2 \over \pi^2} \sqrt{1-\lambda^2}
\int_{-1}^1 d\mu {1 \over \lambda - \mu} {1 \over \sqrt{1-\mu^2}}
\sum_{k=1}^m g_{k+1} (\sqrt{A+B \mu}-b_{0})^k \cr
&\rho(\lambda)=-{2 \over \pi^2} \sqrt{A+B \lambda} \sqrt{1-\lambda^2}
\int_{-1}^1 d\mu {1 \over \lambda - \mu} {1 \over \sqrt{1-\mu^2}}
\sum_{k=1}^m {g_{k+1} \over \sqrt{A+B \lambda}} (\sqrt{A+B \mu}-b_{0})^k
\cr}}
for $n=+2$ and $n=-2$, respectively. Since $\rho(\lambda)$ has to be
non-negative it is easy to prove the positivity condition
\eqn\pos{\eqalign{&\int_{-1}^1 d\mu {1 \over \sqrt{1-\mu^2}}
\sum_{k=1}^m g_{k+1} (\sqrt{A+B \mu}-b_{0})^k=0 \;\;\;\;\;(n=+2) \cr
&\int_{-1}^1 d\mu {1 \over \sqrt{1-\mu^2}} {1 \over \sqrt{A+B \mu}}
\sum_{k=1}^m g_{k+1} (\sqrt{A+B \mu}-b_{0})^k=0 \;\;\;\;\;(n=-2) \cr}}
Together with the normalization condition for the density
$\int_{a'}^{b'} d\mu'\rho(\mu')=1$, i.e.
\eqn\norm{{B \over 2} \int_{-1}^{1} d\mu {\rho(\mu) \over \sqrt{A+B\mu}}
=1\;\;\;\;\;(n=\pm2)}
we are thus given two constraints determining the eigenvalue interval
$[a',b']$ as a function of the couplings $\{b_{0},g_k\}$. Equations
\sol,\pos,\norm\  constitute the exact solution of the $n=\pm2$ loop
equation in the presence of general multicritical matter. We will
now investigate the singular structure of the solution; specific
examples are presented in the next section. The critical points
are located by setting $b'=0$ \gauivan. At the $g^{th}$ multicritical
point we expect the density to behave for $\lambda' \rightarrow 0$
like
\eqn\sing{\rho(\lambda') \sim \lambda'^g}
In order to locate these points we rewrite \sol for $n=+2$ as
\eqn\solplu{\eqalign{&\rho(\lambda)={2 \over \pi} \sqrt{1-\lambda^2}
\sum_{i=0}^{[{m-2 \over 2}]} r_{2 i} (A+B\lambda)^i \cr
&-{2 \over \pi^2} \sqrt{1-\lambda^2} \sum_{l=0}^{[{m-1 \over 2}]}
t_{2 l+1} (A+B \lambda)^l \int_{-1}^1 d\mu {1 \over \lambda - \mu}
{\sqrt{A+B\mu} \over \sqrt{1-\mu^2}} \cr}}
and for $n=-2$ as
\eqn\solmin{\eqalign{&\rho(\lambda)={2 \over \pi} \sqrt{1-\lambda^2}
\sum_{i=0}^{[{m-3 \over 2}]} r_{2 i+1} (A+B\lambda)^{i+{1 \over 2}} \cr
&-{2 \over \pi^2} \sqrt{1-\lambda^2} t_{0} \sqrt{A+B \lambda}
\int_{-1}^1 d\mu {1 \over \lambda - \mu}
{1 \over \sqrt{1-\mu^2}} {1 \over \sqrt{A+B\mu}}\cr
&-{2 \over \pi^2} \sqrt{1-\lambda^2} \sum_{l=0}^{[{m-2 \over 2}]}
t_{2 l+2} (A+B \lambda)^{l + {1 \over 2}}
\int_{-1}^1 d\mu {1 \over \lambda - \mu}
{\sqrt{A+B \mu} \over \sqrt{1-\mu^2}} \cr}}
Here the $\{r_{2i},t_{2l+1}\}$ and
$\{r_{2i+1},t_{2l}\}$ are known functions of the couplings
$\{b_0,g_k\}$ and the parameters $A,B$
(but not of $\lambda$ !) whose precise form we do
not need at present. The remaining integrals in \solplu,\solmin
may be expressed
in terms of complete elliptic integrals of the third kind.
At the critical point $b'=0$
they become elementary. Transforming back to $\lambda'$ \solplu\ may
then be written
\eqn\solpluc{\rho(\lambda')={4 \over \pi a'^2} \sqrt{a'^2-\lambda'^2}
\sum_{i=0}^{[{m-2 \over 2}]} r_{2i} \lambda'^{2i+1}
-{4 \over \pi^2} \sum_{l=0}^{[{m-1 \over 2}]} t_{2l+1} \lambda'^{2l+1}
\log {\lambda' \over a'+\sqrt{a'^2-\lambda'^2}}}
In view of \sing, the possible critical behavior for $n=+2$ is now
transparent from
\solpluc . We note that it is immediately obvious that we can reach only
points with $g$ odd, as expected. There are two kinds of critical
points:

1. $m=g$ (we approach the $g^{th}$ point from
{\it below}.)
It is obtained through the conditions
\eqn\locoplu{\eqalign{&r_{0}=r_{2}=\ldots=r_{g-3}=0 \cr
&t_{1}=t_{3}=\ldots=t_{g-2}=0\;\;\;\;\;t_{g}\neq 0 \cr}}
There are $g-1$ couplings $\{g_{k}\}$ and $g-1$ constraints, so we
indeed have a critical {\it point}.

2. $m=g+1$ (we approach the $g^{th}$ point from {\it above}.) The
conditions are
\eqn\hicoplu{\eqalign{&r_{0}=r_{2}=\dots=r_{g-3}=0\;\;\;\;\;
r_{g-1} \neq 0 \cr
&t_{1}=t_{3}=\ldots=t_{g-2}=t_{g}=0 \cr}}
Here one has $g$ couplings $\{g_{k}\}$ and $g$ constraints; \hicoplu
defines again a critical point. However, one may imagine relaxing the
last condition to $t_{g} \neq 0$ and still be consistent with the
scaling law \sing, see below.

An analogous analysis shows that for $n=-2$ the points with even
$g$ are generated. At $b'=0$ \solmin\  gives

\eqn\solminc{\rho(\lambda')={4 \over \pi a'^2} \sqrt{a'^2-\lambda'^2}
\sum_{i=0}^{[{m-3 \over 2}]} r_{2i+1} \lambda'^{2i+2}
-{4 \over \pi^2} \sum_{l=0}^{[{m-2 \over 2}]} t_{2l+2} \lambda'^{2l+2}
\log {\lambda' \over a'+\sqrt{a'^2-\lambda'^2}}}

The first integral in \solmin is actually divergent
for $b'\rightarrow 0$. This simply gives the condition
$t_{0}=0$. Again one finds two distinct kinds of critical points:

1. $m=g$ (we approach the $g^{th}$ point from
{\it below}.)
It is obtained through the conditions
\eqn\locomin{\eqalign{&r_{1}=r_{3}=\ldots=r_{g-3}=0 \cr
&t_{0}=t_{2}=\ldots=t_{g-2}=0\;\;\;\;\;t_{g}\neq 0 \cr}}

2. $m=g+1$ (we approach the $g^{th}$ point from {\it above}.) The
conditions are
\eqn\hicomin{\eqalign{&r_{1}=r_{3}=\dots=r_{g-3}=0\;\;\;\;\;
r_{g-1} \neq 0 \cr
&t_{0}=t_{2}=\ldots=t_{g-2}=t_{g}=0 \cr}}
Here also one has as many constraints as couplings, but it is consistent
to relax $t_{g}=0$ in 2..

Having found the critical points we will now scale the density in the
vicinity of these points: Introduce a cutoff ${\bar a}$ and define
${\lambda' \over a'}={\bar a} \zeta$ in addition to the usual scaled
separation of the cuts $k'={b' \over a'}={\bar a} M$.
First concentrate on the points
of type 2. Approaching them along a line in coupling constant space
where the conditions \hicoplu, \hicomin remain satisfied we obtain
from \solplu, \solmin for the singular limit of
$\rho(\lambda') \sim {\bar a}^g \rho(\zeta)$
\eqn\singdens{\rho(\zeta)=\zeta^{g-1} \sqrt{\zeta^2-M^2}}
${\bar a}^g w(z)$ being defined as the singular part of
$W(P)=W(P')=-{1 \over 2} \int_{b'}^{a'}d\mu'
{\rho(\mu') \over P'-\mu'}$ (here $P'={1 \over 2}(P_{0}-P)$,
${P' \over a'}=-{\bar a}z$) we derive from \singdens
\eqn\singloop{w(z)=(-1)^{g-1} z^{g-1} \sqrt{z^2-M^2}
\log{{z+\sqrt{z^2-M^2} \over M}}}
Some divergent contributions analytic in $z$, $M$ were discarded in the
course of the derivation; the general meaning of such terms was elucidated in
\mss. Eq.\singloop is already almost \wzn. For $g=1,2$ they in fact
exactly coincide. (Overall wavefunction normalizations are ignored.)
For $g\geq3$ there are additional terms of the form
$z^i M^{g-1-i} \sqrt{z^2 - M^2} \log \ldots$ in \wzn.
In view of eqs.\solplu, \solmin
it is clear how to generate these terms. Approaching the critical
point we have to {\it tune} the $\{g_{k}\}$ such that the
$\{r_{2i}\}$, $\{r_{2i+1}\}$ vanish as the appropriate power of $k'$.
This constitutes, for $g$ integer, the analog of the ``analytic
redefinitions'' of \mss in the half-integer case. Only along special
trajectories in coupling constant space do we satisfy the
Wheeler-de-Witt constraint.

It remains to analyze the relation between the cosmological
constant $\Lambda$ and the scaling parameter $M$.
This is done by investigating
the conditions \pos, \norm. The result is\foot{We have not
carried out a detailed proof except for $g=1,2,3,4$ (see examples)
but strongly suspect the validity of our claim for all $g$.}
\eqn\scal{\eqalign{&\Lambda=M^2 \log {1 \over {\bar a} M}\;\;\;\;\;(g=1) \cr
&\Lambda=M^2\;\;\;\;\;(g\geq2) \cr}}
The critical points of type 1 are essentially different from the type 2
just discussed. They exhibit logarithmic behavior in the eigenvalue
density; it is easily seen that upon scaling we {\it do not} obtain
\wzn and \scal does not hold.

The general theory of this section will now be applied to the lowest
(and most interesting) values of $g$.

\subsec{$g=1$}

This is the loop gas model at $C=1$. $m=1$ is the right boundary of the
dense phase discussed in section 2; $t_{1}=1$. The critical coupling
is $b_{*}=2$. If we introduce the cosmological constant as
$b_{0}^2=2+32 {\bar a}^2 \Lambda$ the relation to the scaling parameter
is $\Lambda=(M \log {\bar a} M)^2$. The density, however, contains
a logarithmic piece, as explained above.

On the other hand,
$m=2$ constitutes the left boundary of the dilute phase. The parameters
in this case read $t_{1}=1-2 g_{1} b_{0}$ and $r_{0}=g_{1} B$. Imposing
the condition \hicoplu, i.e. $g_{1}={1 \over 2 b_{0}}$, the logarithmic
piece in the density is killed and one obtains
\eqn\sole{\rho(\lambda)={1 \over \pi b_{0}} \sqrt{(a'^2-\lambda'^2)(\lambda'^2
-b'^2)}}
Working out the positivity condition \pos  gives $A=b_{0}^2$, while the
normalization constraint yields
\eqn\norme{{2 \over 3 \pi b_{0}}[A a' E(k') - (A^2-B^2) {1 \over a'}
K(k')]=1}
$K(k')$, $E(k')$ are the standard complete elliptic integrals of
the first and second kind, respectively. Considering the limit
$k'\rightarrow 0$ of \norme one locates the critical coupling
to be $b_{0}^2={3 \pi \over 2 \sqrt{2}}$; more importantly, setting
$b_{0}^2={3 \pi \over 2 \sqrt{2}}(1+{\bar a}^2 \Lambda)$ one finds
${\bar a}^2 \Lambda={3 \over 2} k'^2 \log {4 \over k'}$, hence
proving the first assertion in \scal. Note that the Boltzmann weights
of the model are all {\it positive}, as is expected for a unitary
theory\foot{Be aware of our redefinition of the couplings $\{g_{k}\}$
following eq. \sp.}.

\subsec{$g=2$}

Here the central charge is $C=-2$. $m=2$ is the right boundary of the
dilute phase. The parameters are
\foot{Note that for $g_{1}=0$
we are in the dense phase ($m=1$).
The ``critical limit'' thus corresponds to
$b_{0} \rightarrow 0$, meaning infinite Boltzmann weights for loops.
In other words, it does not correspond to a critical theory at all; the
lattice version of ``$C=-\infty$''.}
$t_{0}=-b_{0}+g_{1} b_{0}^2$ and
$t_{2}=g_{1}$. Imposing \locomin gives $g_{1}={1 \over b_{0}}$. The
density contains a logarithmic term. The positivity condition gives
$2 a' E(k')=\pi b_{0}$ while the normalization condition turns
out to be identical to \norme. One thus locates the critical point
to be at $b_{0}^2={24 \over \pi^2}$ and finds
${\bar a}^2 \Lambda ={5 \over 2} k'^2 \log {4 \over k'}$.
Therefore the Boltzmann weights are positive
(a necessary consistency check since we are approaching this point
from the dilute phase); however, there exists a logarithmic scaling
violation $\Lambda=M^2 \log {1 \over {\bar a} M}$.

$m=3$ corresponds to the left boundary of the
first higher multicritical phase (the Yang-Lee phase). One has
$t_{0}=-b_{0}+g_{1} b_{0}^2 -g_{2} b_{0}^3$, $t_{2}=g_{1}-3 g_{2} b_{0}$
and $r_{1}=B g_{2}$. Imposing $t_{0}=t_{2}=0$ it follows
\eqn\solz{\rho(\lambda)={1 \over \pi b_{0}^2} \lambda'
\sqrt{(a'^2-\lambda'^2)(\lambda'^2-b'^2)}}
Positivity gives $A=b_{0}^2$ and normalization $B=2 b_{0}$. Thus
\eqn\kprime{k'=\sqrt{{b_{0}-2 \over b_{0}+2}}}
and it is obvious ($b_{*}=2$) that $\Lambda=M^2$. The
couplings are $g_{1}={3 \over 2 b_{0}}$, $g_{2}={1 \over 2 b_{0}^2}$.
This translates into positive weights for the $\varphi^3$ vertices
and negative weights for the $\varphi^4$ vertices, as it must be
in the Yang-Lee phase. Surprisingly, the model is {\it exactly}
identical to the ``$D=-2$'' theory solved several years ago \KM.
It possesses a Parisi-Sourlas supersymmetry and is most elegantly
described by a zero dimensional supersymmetric matrix model.
The supermatrix is $\Sigma=M+{\bar \theta} \Psi + \Psi^{+} \theta
+{\bar \theta} \theta A$ where $\theta$,${\bar \theta}$ are
Grassmann variables, $\Psi$,$\Psi^{+}$ are Grassmann valued hermitian
matrices and $A$ is the auxiliary hermitian matrix. The action
is\foot{One may equally well use a Gaussian propagator.}
\eqn\susy{
S=N {\Tr} \int d\theta d{\bar \theta}
[{1 \over 2}({\partial \over \partial \theta}\Sigma)
({\partial \over \partial {\bar \theta}}\Sigma)
+{1 \over 2}\Sigma^2 - {g \over 3}\Sigma^3]
}
Upon performing the Grassmann integrals and
integrating out the auxiliary field one obtains
\eqn\effact{
S=N {\Tr}[{1 \over 2}(M-g M^2)^2- \Psi^{+} \Psi(1-2 g M)]
}
By looking at the diagrams the action $S$ is generating one finds the
graphical expansion
\eqn\susygrph{
\sum_{\{\varphi^3 {\rm graphs}\}} g^{v} \sum_{\{{\rm
dimers\;\;+\;\;loops}\}} 3^{v_{0}} (-1)^{{\rm \# (dimers)}}
(-2)^{{\rm \# (loops)}}
}
where $v$ is the total number of vertices and $v_{0}$ the number of
vertices not occupied by either loops or dimers\foot{Because of the zero mode
of the discrete Laplacian \susygrph is of course identically zero for
closed string diagrams. Inserting a loop operator ${1 \over N}\Tr e^{LM}$
fixes the zero mode and yields a nontrivial result.}.
The dimers and loops
are totally self-avoiding (see \loopdim). Now, a comparison of our loop gas
Boltzmann weights with \susygrph immediately reveals the identity of the
two models (the two lattice cosmological constants being related through
$g={1 \over 4 b_{0}}$). It is interesting to note that the loops
are generated by the ``fermionic fields'' $\Psi$,${\bar \Psi}$ while
the dimers are associated with the auxiliary field $A$. What we have
shown here is that the latter degrees of freedom are absolutely
necessary in order to preserve the supersymmetry: without them, the
critical behavior is different.

\subsec{$g\geq 3$}

We will only briefly comment on the type 2 critical point
$m=4$, a $g=3$ model. The conditions that turn off the logarithmic terms are
$t_{1}=1-2 g_{1} b_{0}+ 3 g_{2} b_{0}^2- 4 g_{3} b_{0}^3=0$ and
$t_{3}=g_{2}-4 g_{3} b_{0}=0$. The third parameter is
$r_{0}=g_{1}-3 g_{2} b_{0} + 6 g_{3} b_{0}^2 +g_{3} A$ but now
we have to remember that $r_{0}=0$ upon approaching criticality gives
\singloop while we produce \wzn by {\it tending} $r_{0}$ to zero
appropriately. It is possible if somewhat tedious to prove
${\bar a}^2 \Lambda={5 \over 4} k'^2$ and therefore confirm \scal.

We also carefully investigated $g=4$; no new elements or surprises
are found.

\bigbreak\bigskip\bigskip\centerline{{\bf Acknowledgements}}\nobreak
We would like to thank M.Douglas, E.Martinec, G.Moore, H.Saleur,
N.Seiberg and S.Zamolodchikov for discussions. This work was supported
in part by DOE grant DE-FG05-90ER40559.

\listrefs

\listfigs

\bye